\documentclass[aps,pra,nofootinbib,preprint,amsmath,amssymb,floatfix]{revtex4}
\usepackage{color}
\usepackage{graphicx}

\begin{document}

\newcommand{\tc}{\textcolor}
\newcommand{\g}{blue}
\newcommand{\ve}{\varepsilon}
\title{  Electric Current and Heat Production by a Neutral Carrier: An Effect of the Axion\emph{}}         

\author{ Iver Brevik$^1$ and Masud Chaichian$^2$}      

\affiliation{$^1$Department of Energy and Process Engineering, Norwegian University of Science and Technology, N-7491 Trondheim, Norway}
\affiliation{$^2$Department of Physics, University of Helsinki, P.O.Box 64, FI-00014 Helsinki, Finland}

\date{\today}          

\begin{abstract}

A  general axion-electrodynamic formalism is presented on the phenomenological level   when the environment is dielectric (permittivity and permeability assumed to be constants). Thereafter,
 a  strong and uniform magnetic field is considered in the $z$ direction,  the field region having  the form of a long material cylinder (which corresponds to the haloscope setup). If the  axion amplitude   depends on time only, the axions   give rise to an oscillating electric current  in the $z$ direction. We estimate  the magnitudes of the azimuthal magnetic fields and the accompanying Joule heating in the cylinder, taking the cylinder to have ordinary dissipative properties. We evaluate and calculate the electric current and the heat production separately, without using the effective approximation,  both when there is a strong magnetic field and when there is a strong electric one, showing that with  the magnetic field there is a heat production, while with the electric field there is not.

 The heat generation that we consider, is a nontrivial effect as it is generated by the electrically neutral axions, and has obvious consequences for axion thermodynamics. The heat production can moreover have an additional advantage, since the effect is accumulative and so grows with time. The boundary conditions (in a classical sense) are explained and the use of them  in a quantum mechanical context is discussed. This point is nontrivial, accentuated in particular in connection with the Casimir effect.
  For comparison purposes, we  present finally some results for heat dissipation taken  from the theory of viscous cosmology.

\end{abstract}
\maketitle

\section{Basics}

One of the leading candidates for dark matter in the Universe is the pseudoscalar axion, with amplitude $a=a(x)$, where $x$ means spacetime \cite{preskill83,abbott83,dine83}.
 One believes that the axions are present everywhere, with very weak interaction with ordinary matter, and they are usually taken to move nonrelativistically, with velocities of order $10^{-3}c$.  No strong indication is so far  present regarding their masses, but the common assumption is that the masses lie within an interval of some decades of moderate $\mu$eV/$c^2$. The origin of the axions is connected with processes in the very early Universe, around inflationary times. Their existence was proposed by   Peccei and Quinn in 1977 \cite{peccei77}, in connection with the strong charge-parity (CP) problem in quantum chromodynamics. In turn, the possible existence of these particles has given rise to the so-called axion electrodynamics, a few examples of which are listed in Refs.~\cite{sikivie14,lawson19,asztalos04,sikivie03,mcdonald20,millar17,chaichian20,zyla20,arza20,carenza20,leroy20,brevik20,qiu17,fukushima19,brevik21}.

An important point is whether one is able to detect the axions experimentally, preferably under terrestrial  conditions. It is usual to assume that they are spatially uniformly distributed, thus ${\bf{\nabla}}a=0$, but that they vary harmonically in time as $e^{-i\omega_a t}$, with frequency $\omega_a$. Choosing $m_ac^2= 10~\mu$eV as a reasonable value for the mass, we have from $\hbar \omega_a=m_ac^2$ that $\omega_a=1.52\times 10^{10}~$rad/s, thus a low value, in conformity with the picture of the axions as a classical oscillating field. In principle, an interesting idea is to search experimentally for resonances between the axions and the dielectric  particles in a long plasma cylinder, in the presence of a strong magnetic field in the axial $z$ direction. This is the so-called haloscope approach discussed at various places; cf., for instance, Refs.~\cite{sikivie14,lawson19,asztalos04}. Typical values for resonance frequencies are expected to lie in the region around 100~GHz.

Axion electrodynamics contains many facets, and in the present paper we will be concerned with the oscillating electric current set up in the longitudinal  $z$ direction in a haloscope setup, and the accompanying {\it Joule heating} in the cylinder. We consider both the case when there is a strong magnetic field present, and when there is a strong electric field, with the result that only in the magnetic case will there be a heat production in the cylinder. It is notable that such a heat production is accumulative and so grows with time. The electrodynamic boundary conditions, as taken from classical electrodynamics,  are in these cases nontrivial, and the use of them in a quantum mechanical context is discussed. Actually, these issues are closely related to those appearing in connection with the Casimir effect.

Turn now to the mathematical formalism. The fundamental process is the interaction between a pseudoscalar axion and {\it two} photons \cite{mcdonald20}. The Lagrangian describing the electromagnetic field in interaction with the axion field, in the Heaviside-Lorentz system of units  with $c=1$,
\begin{equation}
{\cal{L}}= -\frac{1}{4}F_{\alpha\beta}{H}^{\alpha\beta}  -\frac{1}{4}g_{a\gamma\gamma}a(x) F_{\alpha\beta}\tilde{F}^{\alpha\beta}. \label{1}
\end{equation}
It is here necessary to include dielectric properties of the surrounding medium. In its rest frame, we will write the constitutive relations as ${\bf D}= \varepsilon{\bf E}, \, {\bf B}=\mu{\bf H}$, with $\bf E$ the electric field and $\bf H$ the magnetic field;  $\bf D$ is the electric induction and $\bf B$ is the magnetic flux density. The material constants are the permittivity $\varepsilon$ and the permeability $\mu$.  As is known, there are two field tensors, the basic tensor  $F_{\alpha\beta}$ and the dielectric response tensor $H_{\alpha\beta}$, where $\alpha$ and $\beta$ run from 0 to 3. We will use the metric convention $g_{00}= -1$. The second term in Eq.~(\ref{1}) should be  a total derivative (topological invariant) when the axion $a=$const, what is obviously true for $F_{\alpha\beta}$ as second factor but not the case with $H_{\alpha\beta}$, for instance.

We give the explicit expressions for the basic field tensor, and the dual of the response tensor,  in our notation,
\begin{equation}
F_{\alpha\beta}= \left( \begin{array}{rrrr}
0    &  -E_x   & -E_y  &  -E_z \\
E_x  &    ~0     & B_z   & -B_y  \\
E_y  &  -B_z   &  ~0    &  B_x \\
E_z  &   B_y   &  -B_x &   ~0
\end{array}
\right). \label{2}
\end{equation}
\begin{equation}
H^{\alpha \beta} = \left( \begin{array}{rrrr}
0    &  -D_x    &  -D_y    &  -D_z  \\
D_x &  0      &  -H_z   &  H_y  \\
D_y &  H_z    &   0     &  H_x  \\
D_z &  -H_y   &   H_x   &    0
\end{array}
\right). \label{2a}
\end{equation}
The definition of the dual  tensor is $\tilde{F}^{\alpha\beta}=\frac{1}{2}\varepsilon^{\alpha\beta\gamma\delta}F_{\gamma \delta}$, with $\varepsilon^{0123}= 1$. The following relations are useful,
\begin{equation}
 F_{\alpha\beta}H^{\alpha\beta}={ 2(\bf{H\cdot B-E\cdot D)}}, \quad  F_{\alpha\beta}\tilde{F}^{\alpha\beta}= -4\,{\bf E\cdot B}. \label{3}
 \end{equation}
In Eq.~(\ref{1}),  $g_{a\gamma\gamma}$ is a combined axion-two-photon coupling constant defined as
\begin{equation}
g_{a\gamma\gamma}= g_\gamma \frac{\alpha}{\pi}\frac{1}{f_a}, \label{4}
\end{equation}
where $g_\gamma$ is another model-dependent constant of order unity. For definiteness, we will adopt the vale $g_\gamma = 0.36$ which follows from the so-called DFS model \cite{dine81}. Further, $\alpha$ is the fine structure constant, and $f_a$ is the axion decay constant whose value is only insufficiently known. One often assumes $f_a \sim 10^{12}~$GeV although it is possible that the value is much lower, around $10^9~$GeV.

We can now write the interaction Lagrangian in the form
\begin{equation}
{\cal{L}}_{a\gamma\gamma} =  g_{a\gamma\gamma} a(x)\,{\bf E\cdot B}. \label{5}
\end{equation}
Based upon the total Lagrangian (\ref{1}), the equations of motion  equations become
\begin{equation}
{\bf \nabla \cdot D}= \rho
-g_{a\gamma\gamma}{\bf B \cdot \nabla}a, \label{6}
\end{equation}
\begin{equation}
{\bf \nabla \times H } = {\bf J}+\dot{\bf D}+ g_{a\gamma\gamma} \dot{a}{\bf B} +g_{a\gamma\gamma}{\bf \nabla}a\times {\bf E}, \label{7}
\end{equation}
\begin{equation}
{\bf \nabla \cdot B}=0, \label{8}
\end{equation}
\begin{equation}
{\bf \nabla \times E}=-\dot{\bf B}. \label{maxwell}
\end{equation}
Here $(\rho, {\bf J})$ are the usual electromagnetic charge and current densities. The dot means time derivative. These equations are in agreement with, for example,  Eqs.~(2.9) in Ref.~\cite{millar17}.

The above  equations can be rewritten as
\begin{equation}
\nabla^2 {\bf E}-\varepsilon\mu \ddot{\bf E}= {\bf\nabla}({\bf \nabla \cdot E})
 +\mu \dot{\bf J}+\mu g_{a\gamma\gamma}\frac{\partial}{\partial t}\left[\dot{a}{\bf B}+ {\bf \nabla}a{\bf \times E}\right], \label{10}
\end{equation}
\begin{equation}
\nabla^2 {\bf H}-\varepsilon\mu \ddot{\bf H}= -{\bf \nabla \times J}-g_{a\gamma\gamma}{\bf \nabla \times }[\dot{a}{\bf B}+{\bf \nabla}a{\bf \times E}], \label{11}
\end{equation}
and further simplifications are achieved  if we omit second order derivatives of the axion, that means time derivatives $\ddot{a}$, space derivatives $\partial_i\partial_j a$, as well as the mixed $\partial_i\dot{a}$ (this has to be checked in the actual situation considered). We then obtain as field equations
\begin{equation}
\nabla^2 {\bf E}-\varepsilon\mu \ddot{\bf E}=  {\bf\nabla}({\bf \nabla \cdot E}) +\mu {\dot{ \bf J}}+  \mu g_{a\gamma\gamma}[ \dot{a}\dot{\bf B} + {\bf \nabla}a{\bf \times \dot{E}}], \label{12}
\end{equation}
\begin{equation}
\nabla^2 {\bf H}-\varepsilon\mu \ddot{\bf H} = -{\bf \nabla \times J}-g_{a\gamma\gamma}
\left[   \dot{a}{\bf \nabla \times B}  + ({\nabla}a){\bf \nabla\cdot E} -[{\bf \nabla}a \cdot {\bf \nabla}]  {\bf E}\right]. \label{13}
\end{equation}
One may here note, using Eq.~(\ref{6}), that
\begin{equation}
{\bf \nabla (\nabla \cdot E)}= \frac{1}{\varepsilon}{\bf\nabla}[ \rho- g_{a\gamma\gamma}{\bf B \cdot \nabla}a]. \label{14}
\end{equation}

\section{Axion-generated Oscillating magnetic fields. Joule heating }
\label{secintro}

In this section we will by means of  simple arguments consider Joule heating problems, both under terrestrial conditions and under astrophysical ones.

\subsection{ Terrestrial considerations}

Assume that there is strong, static and unform magnetic field ${\bf B}_0$ of order 10 T, directed along the $z$ axis. Take the  field region to be a cylinder with radius $R$, and assume for simplicity the cylinder length to be infinite. We will estimate the magnitudes of field strengths, and Joule heating.
In accordance with common assumptions we assume, as mentioned, the axions to be present everywhere, depending on time only. Some recent   references to axion electrodynamics were given above.
 Moreover, we mention the original and interesting work of Caldwell {\it el al.} \cite{caldwell17}  by introducing a dielectric haloscope in the form of  dielectric disks placed in a magnetic field; the new strategy by Lawson {\it et al.} \cite{lawson19} consisting in use of a tunable cryogenic plasma; and the idea of Kim {\it et al.} \cite{kim19} involving  an effective approximation in Maxwell's equations. The last-mentioned authors calculated the {\it difference} between the electric and magnetic stored energies in the cavity, instead of each of these energies  separately.

As mentioned above, we write   the time dependence of the axion as  $a(t)=a_0e^{-i\omega_at}$ with $a_0$ a constant, and  as   the  axion velocity is  so  small $(10^{-3}$),  its frequency  can be put equal to the mass, $\omega_a= m_a$. A value of  $m_a= 10\, \mu$eV seems to be a reasonable choice, given the present uncertainties, so that the axion becomes pictured as a classical field object.

How big is the amplitude  $a_0$?  We may express $a(t)$ in terms of the angle $\theta(t)$ characterizing the QCD vacuum state,
\begin{equation}
a(t)= f_a\theta(t), \label{4a}
\end{equation}
 so that  the axion field  becomes
 real, $a(t)=a_0\cos \omega_at$, and similarly $\theta(t)= \theta_0\cos \omega_at$.  For the amplitude we thus have  $a_0=f_a\theta_0$.  The axion current density in the $z$ direction becomes (we omit the permeability, which is of order one and without practical significance)
\begin{equation}
 J_{\rm axion}(t) =  g_{a\gamma\gamma} \dot{a}(t){ B}_0=   -\left( g_\gamma \frac{\alpha}{\pi} \omega_aB_0\right) \theta_0\sin \omega_at. \label{5a}
\end{equation}
It is worth noticing that  only the ratio  $a_0/f_a=\theta_0$ is involved here.
 Experimental information \cite{harris99}) shows that the value of $\theta_0$ is  small. From Ref.~\cite{graham11} we may quote
 \begin{equation}
 \theta_0 \sim 3\times 10^{-19}, \label{6a}
 \end{equation}
 but we henceforth  regard   $\theta_0$ only as an unspecified  parameter.

 When dealing with numerical estimates it is convenient to use SI units:
\begin{equation}
 J_{\rm axion}(t) =    - \left( g_\gamma \frac{\alpha}{\pi} \omega_a \frac{B_0}{\mu_0c}\right) \theta_0\sin \omega_at.\label{stromtetthet}
\end{equation}
After   multiplying  with the cross sectional area $\pi R^2$ we get   the axial current in the cylinder, $I_{\rm axion}(t)= I_0\sin \omega_at$, with dimension A (amp\`{e}re).

With $g_\gamma = 0.36, \omega_a= 1.52\times 10^{10}~$rad/s,  $B_0= 10~$T,
and by choosing the cylinder radius as $R=10~$cm,  we obtain for the axion current
\begin{equation}
I_{\rm axion}(t) \approx  - 3\times 10^5 \times \theta_0 \sin \omega_at. \label{strom}
\end{equation}
In principle, it is  possible to detect this current via the azimuthal magnetic field $B_\theta(r,t)$ it generates. Assume that the cylindric region of radius $R$ is filled with a nonmagnetic dielectric cylinder, and that there is a vacuum on the outside. As the frequency $\omega_a$ is low, we may make use of the theory under  quasi-static conditions \cite{landau84}. They apply when the frequency of the field is small compared to the inverse mean free time for the microscopic conductivity mechanism and implies, as the polarization current density is neglected,  that intricate boundary condition problems noted in Ref.~\cite{kim19} are avoided. In cylindrical coordinates, on the outside, we then get
\begin{equation}
B_\theta (r>R)= \frac{\mu_0}{2\pi}\frac{I_0}{r} \sin \omega_a t =  -\frac{6}{r}\times 10^{-2}\times \theta_0 \sin \omega_at.  \label{9}
\end{equation}
This is very small. Taking $\theta_0 \sim 10^{-19}$ as mentioned above, we get for the amplitude at $r=R$
\begin{equation}
B_\theta (r=R) \sim 10^{-19}~\rm T.
\end{equation}
 SQUID magnetometers are able to measure magnetic fields down to the $10^{-15}~$T region. Thus, in order to be able to measure the oscillating azimuthal magnetic field by this simple method, $\theta_0$ has to be much larger than $10^{-19}$.

 Let us consider the boundary conditions  at $r=R$ more closely. They are adopted from classical electrodynamics, implying continuity conditions for the tangential components of  $\bf E$ and $\bf H$ at  $r=R$.  This appears nonproblematic in the present case, as we can easily calculate the azimuthal magnetic field in the interior,
 \begin{equation}
 B_\theta(r<R)= \frac{\mu_0}{2\pi}\frac{I_0r}{R^2}\sin \omega_at,
 \end{equation}
 and so check by comparison with Eq.~(\ref{9}) that $B_\theta$ is continuous at $r=R$. However in a  general context the boundary problem  is more delicate, as one is usually applying   a
 {\it classical} quantity such as permittivity in a {\it quantum} field theory. The boundary conditions thus get a hybrid character. \emph{}The issue becomes accentuated in connection with the Casimir effect. A striking example is shown in the Casimir theory of a dielectric ball \cite{milton80}. The Casimir pressure on the surface is calculated using the radial component of the Maxwell stress tensor, in which case the separation distance between two neighboring spacetime points $x$ and $x^\prime$ occurs. In order to handle the formal divergence in the limit  $x \rightarrow x^\prime$, one introduces a time splitting such that the difference $\tau = t-t^\prime$ is small but finite. The important point is that if this cutoff term is kept in the final expression for the pressure, it corresponds to a physical property, namely {\it surface tension}. In this way the mathematical trick of time splitting is related to microscopic physics. Actually, if one inserts  data for surface tension for usual fluids such as water, it turns out that the distance $\tau c$ travelled by light during the time interval $\tau$ is of the same order of magnitude as atomic dimensions,
 \begin{equation}
 \tau c \sim 0.1~\rm{nm}.
 \end{equation}
 These considerations indicate that there is a link between quantum field theory and microscopic statistical mechanics, caused by the use of the classical permittivity concept in the boundary conditions in QFT.
 A more detailed discussion on this issue is given in Ref.~\cite{hoye17}.

 \bigskip

 \noindent {\it Joule heating.~~} Turn now to the low axion-generated generation of heat.  It is natural to  construct  a  "cylindrical-like" object such that its oscillation frequencies are low, in order to match $\omega_a$. This is, as mentioned, the haloscope approach.  There exists  extensive theoretical and experimental works in this direction \cite{asztalos04,sikivie14,caldwell17,lawson19,kim19}. Here,
 we will present only the main idea.

 The heat $Q$ produced per unit length is
\begin{equation}
Q= \sigma E^2\pi R^2 = \frac{I_0^2}{\sigma \pi R^2}\sin^2 \omega_at.
\end{equation}
With $I_0 = - 3\times 10^5\,\theta_0$ from Eq.~(\ref{strom}) we obtain, by choosing
  $R=10~$cm  and  $\sigma = 10^4~$S/m
(a typical value for semiconductors), that $ Q \approx  2\times 10^8\,\theta_0^ 2 \,\sin^2\omega_at$,
with dimension $[Q]=$W/m.
 With $\theta_0= 10^{-19}~$, this value  corresponds to a dissipation per unit volume $q$ equal to
\begin{equation}
q=\frac{Q}{\pi R^2} \sim 10^{-28}\sin^2\omega_at. \label{terrestrial}
\end{equation}
The existence of this dissipation, although small, raises questions of fundamental interest. First, the theory predicts there to be  an electric current, produced by  neutral particles  \cite{wheeler}.  Moreover, from  a thermodynamic viewpoint one may ask: what is the source generating the heat? It cannot be the field $B_0$, because a static magnetic field does not convey heat. So, the energy source must necessarily be the axions themselves. This brings up the question about thermal equilibrium. If the axions  are in a state of thermodynamic decay, and have   existed for a long time, there should have been  several possibilities for them to have disappeared by now. Thermodynamic aspects of the axion physics seem not to be yet well settled. It is likely that this behavior is related to our neglect of backreaction.

\subsection{Astrophysical considerations}

For comparison purposes, since axions are assumed to be present everywhere, it becomes natural to look at astrophysical objects. In particular, magnetars are known to have strong  magnetic fields, in the range of $10^8$ to $10^{11}~$T (tesla).  Let us for definiteness take $B_0= 10^{10}~$T on the surface of a neutron star of 1.5 sun masses with typical radius $R=12$ km. Its volume is $V= (4\pi/3)R^3 = 7.2\times 10^{12}~$m$^{3}$. Assume as a rough model that the star has the form of a cubic box with the same volume $V$, acted upon by this strong field $B_0$ directed in the $z$ direction. An  oscillating electric current will occur in the $z$ direction. Inserting $B_0= 10^{10}~$T in Eq.~(\ref{stromtetthet}) we obtain for the current density in the present case $
J(t)= 1.7\times 10^{14}\times \theta_0 \sin \omega_at$.  In order to calculate the local dissipation of heat, $J^2(t)/\sigma$, we need the electrical conductivity $\sigma$ in a neutron star. This value is  very high; from Ref.~\cite{baym69} we quote $\sigma = 1.7\times 10^{18}~$S/m. Then $
{J^2(t)}/{\sigma}= 1.7\times 10^{10}\times \theta_0^2\sin^2\omega_a t$,
with dimension W/m$^3$. The rate of heat produced in our 'pulsar' is found by multiplying with $V$. Moreover, as the number of pulsars observed in our galaxy is about $N=2000$, we find for the total pulsar-generated dissipation,  $
Q_{\rm pulsars}= \left( J^2(t)/{\sigma}\right)V N. $
Considering our galaxy as a disk with radius $10^{21}~$m and thickness $10^{19}~$m, thus with volume $3\times 10^{61}~$m$^3$, we can calculate the mean dissipation $\bar{q}_{\rm pulsars}$ per unit volume,
\begin{equation}
\bar{q}_{\rm pulsars}= 4\times 10^{-36}\times \theta_0^2, \label{pulsar}
 \end{equation}
with dimension W/m$^3$. This expression is seen to be much smaller than   the expression (\ref{terrestrial}) obtained under extremal terrestrial conditions.

As a second example taken from astrophysics we will consider viscous cosmology. This variant of cosmological theory  implies introduction of viscous terms in the fluid's energy-momentum tensor $T^{\mu\nu}$ (though bulk viscosity $\zeta$ only, in view of isotropic symmetry). We will assume a one-component cosmic fluid, with energy density $\rho$.  Its development is governed by the Friedmann equations, although in our context it is sufficient to consider the energy-conservation equation only,
\begin{equation}
\dot{\rho}=-3H(\rho+p) + 9\zeta H^2,
\end{equation}
where $\zeta$ denotes the bulk viscosity (spatial curvature $k=0$  is assumed), and $H$ is the Hubble parameter. From this equation it follows that the viscous dissipation comes from the last term to the right. Calling this term $q_{\rm viscous}$, we have
\begin{equation}
q_{\rm viscous}   = 9\zeta H^2.
\end{equation}
By comparing measured values of $H(z)$ as function of the redshift $z$ with theoretical predictions for different input values for $\zeta$, it follows that best agreement is obtained when  $\zeta$ is nonvanishing.  Cf., for instance,  Ref.~\cite{wang14} for a study in this direction.   For definiteness we adopt here the following value for $\zeta= \zeta_0$ at present time \cite{brevik16,normann20},
\begin{equation}
\zeta_0= 10^5~\rm {Pa~s}.
\end{equation}
Taking the present-time value of the Hubble parameter to be
 $H_0=67.74~$km$^{-1}$Mpc$^{-1}$= 2.20$\times 10^{-18}~$s$^{-1}$ we  get
 \begin{equation}
 q_{\rm viscous}=9\zeta_0H_0^2 = 4.3\times 10^{-30}, \label{viscous}
 \end{equation}
with dimension W/m$^3$. This is a phenomenological quantity derived from experiments, thus independent of any assumption about $\theta_0$. Its value  is actually not far off the result (\ref{terrestrial}) for extremal terrestrial conditions when $\theta_0 \sim 10^{-19}$.
     As mentioned, the  important property of Eq.~(\ref{viscous}) is that it relies upon the Friedmann equations only, and is thus independent of any assumption about the phase $\theta_0$ of the axion. Again, this may be related to our neglect of backreaction.

\section{Discussion}

\noindent 1. It ought to be emphasized that we have considered axion electrodynamics at the perturbative level; the axion field has been taken to be not a dynamical field. The latter option would imply to consider the detailed motion  of the axions, about which no information is available. That  would make any estimate on the production of heat with a quantitative value for it, impossible. The point is that what is known is only the average density of axions in the space. At the same time, using an average density of axions, this is equivalent to consider the axion field to have no  dependence on time or coordinates in the whole long cylinder. In addition, if one would like to take into account the backreaction of axions, that would take us to the higher orders of perturbation and disable us to make   any numerical estimate for the generated heat, as well as for the majority of other effects in physics too.

\noindent 2. An experimental detection of the axion particles is evidently a demanding task. As mentioned, the  haloscope approach  may be a feasible method \cite{sikivie14,lawson19,asztalos04,caldwell17,kim19},  although one  must  then be able to lower the resonance frequencies $\omega_0$ in a dielectric  cylinder so much that there occurs  approximate coincidence with the axion frequency $\omega_a \approx 10^{10}~$rad/s. There are various ways of doing this \cite{caldwell17,kim19}. For instance, in Ref.~\cite{caldwell17} a special variant of a dielectric haloscope was proposed, aiming to detect axions in the high-mass range $40-400~\mu$eV. A stack of a large number $N$ of parallel plates ($N \sim 80$) was assumed to be situated parallel to a parallel mirror. This setup implied  advantages from a large transverse area,  and also from the possibility of making both broadband and narrow-band tests.

The axion-induced effects, in general, are very small, as elaborated on  in  Sec.~II where we provided some numerical   estimates. The main point is the generation of heat from axion particles being electrically neutral.  Our calculation did not make use of the effective approximation, an essential simplifying factor here being our restriction to low (quasi-static) frequencies, whereby the general boundary condition problem is avoided.
A noteworthy characteristic property of the heat production is that it accumulates with time.

\noindent 3. When dealing with the boundary conditions in Sec.~II.A we assumed the quasi-static conditions, which are applicable when the field frequency is much less than the inverse microscopic free time. Thereby these conditions become simplified. In general, however, if the usual boundary conditions for dielectric surfaces are simply carried over in a QFT context, delicate issues occur which are not well solved at present (the Casimir effect is a striking example). We discussed also this point briefly, in the same section.

\noindent 4. Another  novel  element in our analysis was the comparison with heat generation in cosmological theory, Sec.~II.B. We made here use of the bulk viscosity only, which is an assumption compatible with spatial isotropy of the cosmic fluid. The value $\zeta_0 = 10^5~$ Pa\,s that we made use of, appears to conform fairly well with cosmological observations.

\noindent 5. From Maxwell's equation (\ref{6}) - (\ref{maxwell}) it follows that even by taking a time dependent axion within the entire cylinder, $a=a(t)$, still there is no coupling between axions and  the electric field at all.  Thus the electric and the magnetic fields behave differently. As a result, no heat production will take place.   On a basic level, this breaking of electromagnetic duality is related to the lack of  magnetic charges in Maxwell's equations.

\noindent 6. Finally, we may consider again the following question:  where does the produced heat come from? Recalling  the discussion on this point at the end of Sec.~II.B, we may summarize and extend: The strong magnetic field $B_0$ can not produce heat. Therefore, the probable source for the heat generation is the axion fluid itself. In turn, this raises however the natural follow-up question: why have not the axions, which  have existed for a long time in the universe, already transferred away all excess heat? This is a delicate problem that reduces in the end to basic thermodynamics, and is in our opinion not solved so far. Most likely, this kind of behavior is related to our perturbative description of the axion field.

Comparison with the  cosmological viscous approach is here helpful. We calculated the heat generation using the classical viscosity concepts, thus without any bearing on the axion field at all. The use of the two viscosity coefficients (shear, and bulk) in viscous fluid dynamics means effectively  that one is working to the first order in deviations from thermal equilibrium. The assumption about spatial isotropy in the Universe means that only the bulk viscosity remains actual. So, it is clear that the physical reason for cosmological produced heat can be attributed to fundamental thermodynamics with a bulk viscosity present. It is intriguing to wonder if not the same physical property lies at the bottom of the heat generation from axions also, thus within the realm of quantum mechanics.


\section*{Acknowledgements}

We are deeply grateful to  Kimball A. Milton and Yuri N. Obukhov for several  valuable remarks and suggestions on the manuscript.

\end{document}